\documentclass[a4paper]{article}
\usepackage{hyperref}
\usepackage{tikz}
\usepackage{pgfplots}
\usepackage[skip=5pt]{caption}
\usepackage{lineno}

\newcommand{\etal}{\textit{et al}.}
\newcommand{\figcap}{\setlength{\baselineskip}{.5em}}

\usepackage{INTERSPEECH2018}

\title{Nebula: F0 Estimation and Voicing Detection by Modeling the Statistical Properties of Feature Extractors}
\name{Kanru Hua}
\address{University of Illinois, U.S.A.}
\email{khua5@illinois.edu}

\begin{document}

\maketitle

\begin{abstract}

A F0 and voicing status estimation algorithm for high quality speech analysis/synthesis is proposed. This problem is approached from a different perspective that models the behavior of feature extractors under noise, instead of directly modeling speech signals. Under time-frequency locality assumptions, the joint distribution of extracted features and target F0 can be characterized by training a bank of Gaussian mixture models (GMM) on artificial data generated from Monte-Carlo simulations. The trained GMMs can then be used to generate a set of conditional distributions on the predicted F0, which are then combined and post-processed by Viterbi algorithm to give a final F0 trajectory. Evaluation on CSTR and CMU Arctic speech databases shows that the proposed method, trained on fully synthetic data, achieves lower gross error rates than state-of-the-art methods.

\end{abstract}

\section{Introduction}

The problem of estimating the fundamental frequency (F0) and voicing status of speech signals has been extensively explored using a combination of signal processing and heuristic techniques. Classical methods rely on time-domain measurement of auto-correlation \cite{talkin-1995} or normalized auto-correlation \cite{decheveigne-2002}. Selection of F0 candidates in spectral domain \cite{drugman-2011} and mixed domain \cite{morise-2010} also has been studied with varying degrees of consistency across databases and noise levels.

We see a recent trend in the rise of probabilistic F0 estimation methods, often as an attempt to reduce the use of heuristic elements in the algorithm and ultimately to achieve more consistent performance without expert's intervention. In particular, a class of data-driven methods indirectly perform F0 estimation by doing inference on features extracted from the input signal. Notably, SAFE \cite{chu-2012} (Statistical Approach to F0 Estimation) bears similarities to our method in that a statistical framework is employed in which signal-to-noise ratio (SNR) features are used to aid the discrimination of harmonic against noise. However, the method specifically designed for information extraction from SNR peaks does not allow for incorporating other types of signal features. Another related approach is SAcC \cite{lee-2012} (Subband Auto-correlation Classification), which predicts the distribution of F0 using a feed-forward neural network trained on frequency-dependent auto-correlation functions.

Modeling speech features instead of formulating the problem directly on the waveform makes the model less prone to inaccurate assumptions on speech signals, aside from reducing the mathematical complexity. The downside is that characterization of speech features often relies on data-driven techniques such as distribution fitting and regression, making the performance data-dependent to some extent. YANG \cite{kawahara-2016} (Yet ANother Glottal source analysis framework), a more recent method finds a balance between the use of heuristics, probabilities and data-dependent parameters. YANG first divides the input speech into overlapping frequency channels. For each channel, SNR and instantaneous frequency features are extracted at a fixed time interval. The features from all channels are converted into a mixture distribution on F0 via a set of heuristics and a smooth F0 trajectory is tracked using a Viterbi search. Our previous work \cite{hua-2017} successfully reduced the fine error of YANG algorithm by calibrating the SNR estimator on synthetic speech data. This study takes the idea of data-free modeling of speech feature extractors a step further by training Gaussian mixture models (GMM) on the entire feature extraction framework with synthetic speech as the input. We show that good generalization can be achieved with the appropriate choice of feature extractors meeting certain assumptions allowing the relaxation of the synthetic data generator.

\textbf{This paper is organized as follows:} Section 2 begins with some theoretical discussion that sets the ground for the algorithm design, followed by an overview of the proposed method. The F0 estimation and voicing detection stages are explained in Section 3 and Section 4 respectively. Section 5 evaluates the proposed method on two speech databases and analyzes the results. Finally, this paper is concluded in Section 6.

\section{Overview}
\label{sec:overview}

While the strategy of training regression or classification models on synthetic data has received moderate attention in image recognition \cite{varga-2003, jaderberg-2014}, to our knowledge the idea is rather under-explored in the area of speech analysis, possibly due to the lack of a high-quality speech synthesizer that matches the distribution of natural speech signals. We circumvent the chicken-and-egg problem of building a near-perfect speech synthesizer for studying speech analysis by asking, under what conditions can the requirements on training data be relaxed? This question points us to the general methodology of problem reduction: in the context of speech signal analysis, factorizing the scope and dimension of variables being modeled down to the vicinity of a time-frequency point. The time-frequency localized problem only requires a synthetic data generator that reproduces fragments of speech signals at a microscopic level, for example the sum of short sine waves and noises, without modeling the formant structure or F0 variations.

\begin{figure}
  \centerline{\includegraphics[width=7cm]{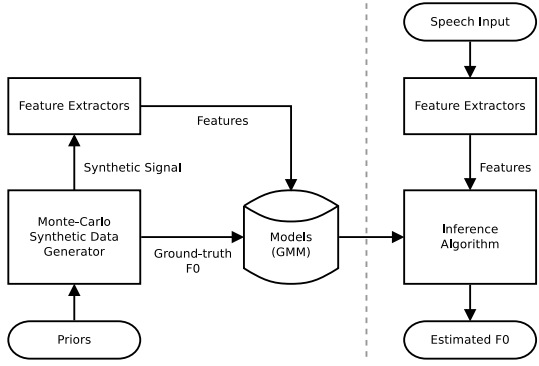}}
  \caption{\figcap{An overview of Nebula, the proposed training-data-free F0 estimation framework with training and inference phases separated by the dashed line.}}
  \label{fig:overview}
\end{figure}

The said locality condition can be easily met using a short analysis window and band-pass filtering; the reduction of time-domain waveforms to a small number of variables can be done using a set of feature extractors. It is found that YANG \cite{kawahara-2016} provides a powerful feature extraction framework satisfying all of the conditions mentioned. Specifically, at each frame the SNR and instantaneous frequency (IF) are estimated from a set of logarithmically spaced overlapping frequency channels covering the first few harmonics. Thus a distribution over SNR, IF and F0 can be defined for every time-frequency point. As outlined in Figure~\ref{fig:overview}, the distributions are found by fitting Gaussian mixture models (GMM) on synthetic data generated from Monte-Carlo simulations. Once the models are trained, the conditional distribution on F0 can be computed from arbitrary input data, as long as the time-frequency local distributions of the unseen data are covered by the priors for Monte-Carlo simulations.

\begin{figure}
  \centerline{\includegraphics[width=7.8cm]{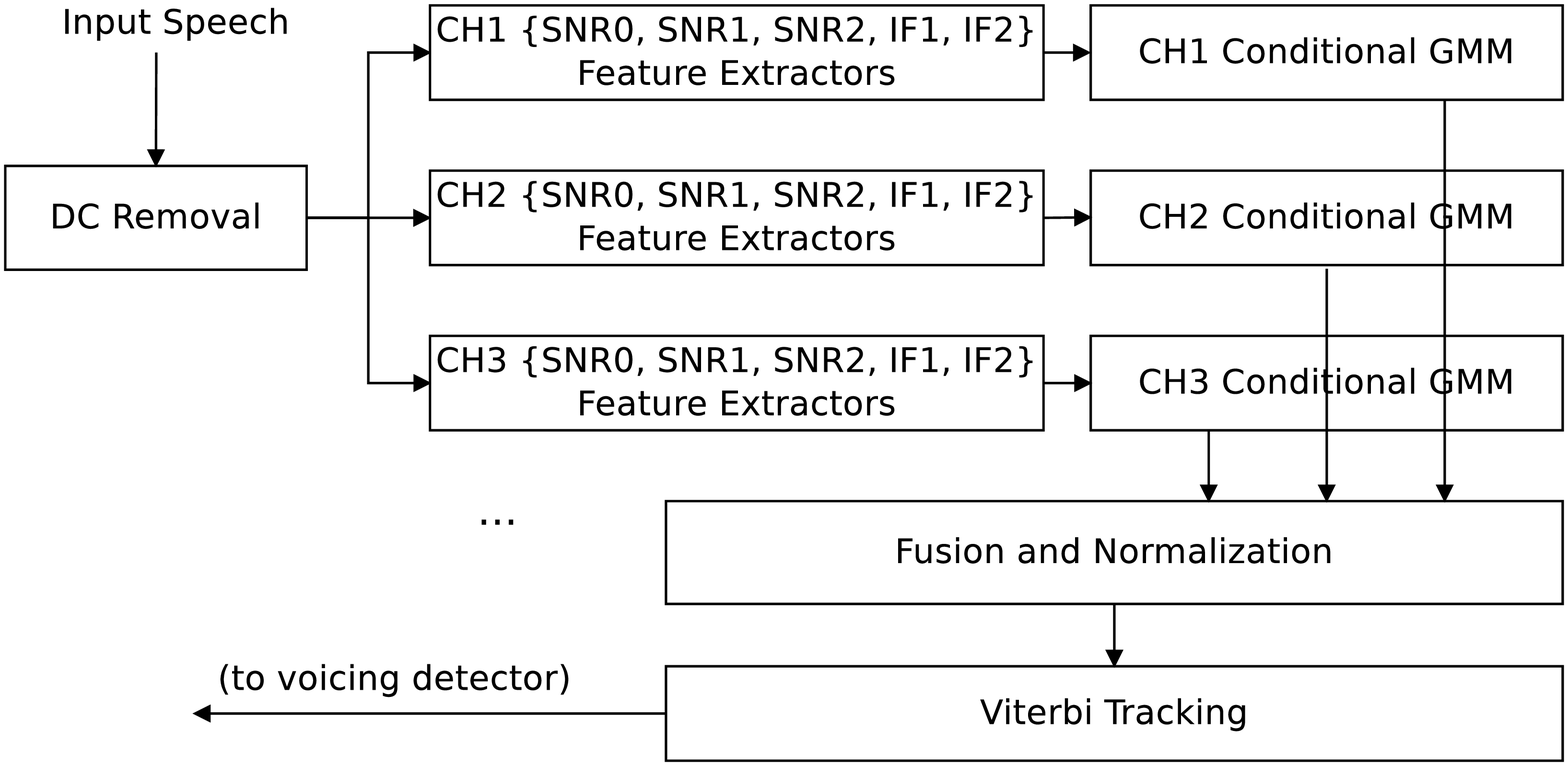}}
  \caption{\figcap{Flowchart of the F0 inference subroutine.}}
  \label{fig:f0-flowchart}
  \vspace{-10pt}
\end{figure}
\subsection{Signal Model for the Data Generator}
\label{sec:priors}

As seen from the above discussion, the proposed system is designed to model the statistical properties of speech feature extractors instead of the complicated process of speech itself. The rationale for choosing the priors for data fabrication, rather than mimicking speech signals, becomes covering as much of the assumption-defined signal space as possible to fully characterize the feature extractors. Though being less relevant, the expert knowledge on speech phenomena is specified implicitly through the choice of feature extractors and the signal model for the data generator. In this study the synthetic data is generated from a harmonic-noise model defined as the follows,
\begin{align}
  x[n] &= \mathbf{a_t} u[n] + \sum_{k=1}^K \mathbf{a}_k \sin(2\pi n k \mathbf{f_0} / f_s + \theta_k) \\
  u[n] &\sim \mathcal N(0, 1),\quad \theta_k \sim \mathcal U(-\pi, \pi) \nonumber
\end{align}
Shown in boldface are the random variables specified by the priors: $\mathbf{a_t}$ is the overall SNR following a log-uniform distribution in $[-50, 50]$ dB; $\mathbf{a}_k$ is the amplitude of the k-th harmonic following a log-uniform distribution in $[-10, 10]$ dB; $\mathbf{f_0}$ is the fundamental frequency following a log-uniform distribution in $[40, 1000]$ Hz covering both speech and singing.

The widespread use of log-uniform priors covers a significant portion of the feature space and ideally should improve the generalization across speakers. For such the reason the proposed algorithm is named Nebula. The inference part of the algorithm is described in the following sections.

\section{Conditional GMM based F0 Estimation}

Figure~\ref{fig:f0-flowchart} outlines the F0 inference method in Nebula; the flowchart elaborates the right side of Figure~\ref{fig:overview}. The input speech, after removing the DC component, is processed by a filterbank with 36 sets of feature extractors. Aside from the SNR and IF estimators featured in the original YANG \cite{kawahara-2016} algorithm, for each channel a second set of estimators (``SNR2" and ``IF2") are added at twice the channel frequency and a third SNR estimator (``SNR0") is added at half the channel frequency. Although the inclusion of feature extractors at different frequencies violates the frequency locality condition (section~\ref{sec:overview}), our preliminary test revealed a reduction in double and half frequency errors that leads to a lower overall error rate.

\begin{figure}
\begin{minipage}[b]{1.0\linewidth}
%
%
\begin{tikzpicture}

\begin{axis}[%
width=2.6in,
height=0.8in,
at={(0.953in,0.294in)},
scale only axis,
axis on top,
xmin=0,
xmax=1675,
ymin=0,
ymax=128,
ylabel near ticks,
xtick distance=300,
ylabel={frequency index},
axis background/.style={fill=white}
]
\addplot [color=blue, mark=*, mark options={solid, blue}, forget plot]
  table[row sep=crcr]{%
1	-100\\
};
\addplot [forget plot] graphics [xmin=0.5, xmax=1675.5, ymin=0.5, ymax=128.5] {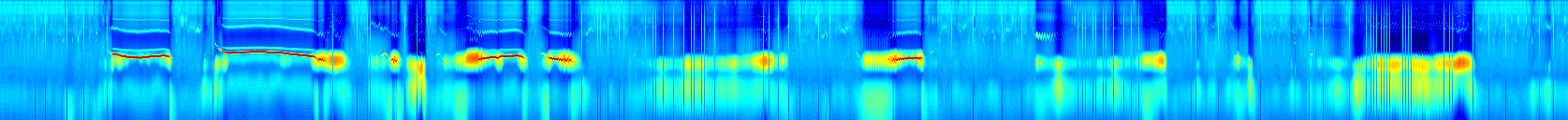};
\end{axis}
\end{tikzpicture}%
\end{minipage}
\begin{minipage}[b]{1.0\linewidth}
\input{fig-Lmap}
\end{minipage}
\begin{minipage}[b]{1.0\linewidth}
\input{fig-vud}
\end{minipage}
\caption{\figcap{From top to bottom: $\log p_{20}(f | \mathbf{x}_{20})$ computed from a speech sample; estimated F0 trajectory superimposed on the log likelihood map; F0 likelihood and log voicing probability along the F0 trajectory.}}
\label{fig:multi-plot}
\vspace{-10pt}
\end{figure}

The rest of the inference algorithm focuses on converting feature vectors into posterior distributions on F0, and performing tracking on a likelihood map combining the posterior distributions from all frames. For the k-th channel, we first define feature vector $\mathbf{x}_k$ to be,
\begin{align}
  \mathbf{x}_k &= [\mathbf{SNR0}_k, \mathbf{SNR1}_k, \mathbf{SNR2}_k, \mathbf{IF1}_k, \mathbf{IF2}_k]^T
\end{align}
Each GMM models the joint distribution over F0 and feature vectors. The feature vector augmented by the random variable on F0 is denoted as $\mathbf{y}_k$,
\begin{align}
  \mathbf{y}_k &= [\mathbf{x}^T_k, \mathbf{f_0}_k]^T
\end{align}
Our interest lies in recovering the augmented vector $\mathbf{y}_k$ from its truncated version $\mathbf{x}_k$, which is essentially to estimate the last element $\mathbf{f_0}_k$. Then the estimates from multiple channels can be combined to give a more robust posterior distribution. Given the GMM trained on each channel using the synthetic data defined in section~\ref{sec:priors}, the recovery of $\mathbf{f_0}_k$ from $\mathbf{x}_k$ is done in a way similar to GMM-based voice conversion \cite{kain-1998}. Concretely, the joint density for the k-th channel is defined as,
\begin{align}
  p_k(\mathbf{y}_k) &= \sum_m w_{mk} \mathcal N \left( \mathbf{y}_k | \mu_{mk}, \Sigma_{mk} \right) \\
  \Sigma_{mk} &= \begin{bmatrix}
      \Sigma^{\mathbf{x}}_{mk} & \Sigma^{\mathbf{x} \mathbf{f_0}}_{mk} \\
      \Sigma^{\mathbf{f_0} \mathbf{x}}_{mk} & \sigma^{\mathbf{f_0}}_{mk}
    \end{bmatrix},
  \mu_{mk} = \begin{bmatrix}
      \mu^{\mathbf{x}}_{mk} \\ \mu^{\mathbf{f_0}}_{mk}
    \end{bmatrix} \nonumber
\end{align}
Given a feature vector $\mathbf{x}_k$ built from the extracted features, the GMM over $\mathbf{y}_k$ is converted into a single-dimensional GMM over the conditional distribution $\mathbf{f_0}_k | \mathbf{x}_k$, an example of which is shown in the upper plot of Figure~\ref{fig:multi-plot},
\begin{align}
  p_k(\mathbf{f_0}_k | \mathbf{x}_k) &= \sum_m w'_{mk} \mathcal N \left( \mathbf{f_0}_k | \mu'_{mk}, \sigma'_{mk} \right) \\
  w'_{mk} &= \frac{w_{mk} \mathcal N (\mathbf{x}_k | \mu^\mathbf{x}_{mk}, \Sigma^\mathbf{x}_{mk} )}
    {\sum_n w_{nk} \mathcal N (\mathbf{x}_k | \mu^\mathbf{x}_{nk}, \Sigma^\mathbf{x}_{nk} )} \nonumber \\
  \mu'_{mk} &= \mu^\mathbf{f_0}_{mk} + \Sigma^{\mathbf{f_0} \mathbf{x}}_{mk} \Sigma^{\mathbf{x}^{-1}}_{mk} (\mathbf{x}_k - \mu^\mathbf{x}_{mk}) \nonumber \\
  \sigma'_{mk} &= \sigma^\mathbf{f_0}_{mk} - \Sigma^{\mathbf{f_0} \mathbf{x}}_{mk} \Sigma^{\mathbf{x}^{-1}}_{mk} \Sigma^{\mathbf{x} \mathbf{f_0}}_{mk} \nonumber
\end{align}

Next, the conditional probabilities from all channels are combined under an independence assumption. However due to the correlation between features in neighboring channels, a simple summation of log conditionals would over-emphasize the modes. This problem can be easily addressed by taking the average of log conditionals instead. The result is an unnormalized log likelihood, denoted as $\mathcal L^-$.
\begin{align}
  \mathcal L^-(f | \mathbf{x}_{1,2,..., K}) &= \frac{1}{K} \sum_{k=1}^K \log p_k(f | \mathbf{x}_k)
\end{align}
An important but non-obvious issue regarding the unnormalized likelihood $\mathcal L^-$ is that, due to the non-uniform spacing of frequency channels and the log-uniform distributed priors for the Monte-Carlo simulation (section~\ref{sec:priors}), $\mathcal L^-$ could be biased towards a certain frequency range. Inspection of the results on speech data tells that $\mathcal L^-$ exhibits a systematic bias favoring lower frequencies. This bias can be compensated by subtracting the expectation of unnormalized likelihood computed on white noise inputs from $\mathcal L^-$ during inference. The said expectation is denoted as the calibration function $\mathcal L_\mathrm{cal}$. After normalization, a log posterior density $\mathcal L(f)$ is obtained on each frame,
\begin{align}
  \mathcal L_\mathrm{cal}(f) &= \mathrm{E}_{x[n] \sim \mathcal N(0, 1)}[\mathcal L^-(f | \mathbf{x}_{1,2,..., K})] \\
  \mathcal L(f) &= \mathcal L^-(f | \mathbf{x}_{1,2,...,K}) - \mathcal L_\mathrm{cal}(f) - \\
    &\phantom{=} \log \int \exp[\mathcal L^-(f' | \mathbf{x}_{1,2,...,K}) - \mathcal L_\mathrm{cal}(f')] df' \nonumber
\end{align}
The procedure described above, from feature extraction to computing the log posteriors, is repeated at a fixed time interval, yielding a likelihood map $\mathcal L(f, t)$ across time and frequency (the second plot in Figure~\ref{fig:multi-plot}). To robustly track the peak frequency, the likelihood map is first sampled on a log-spaced frequency grid and then passed into a Viterbi path searcher as the observation probability. The transition probability for the Viterbi search, which constrains the first-order log F0 dynamics, is set according to a zero-mean normal distribution with a standard deviation of $2\ \mathrm{oct/s}$. The resulting sequence of frequency indices is refined using quadratic interpolation on the likelihood map, similar to the quadratically-interpolated FFT method for sinusoidal analysis \cite{smith-1987}. Finally, the F0 estimation algorithm gives a continuous log-F0 trajectory. The voicing status has yet to be determined, as explained in the following section.

\section{Voicing Detection}

The F0 estimation stage outputs a time-frequency F0 likelihood map. Over regions exhibiting strong periodicity, the F0 likelihood tends to be unimodal across frequency; over noisy or silent regions, the likelihood is general flat, as exemplified in the second plot in Figure~\ref{fig:multi-plot}. It thus comes naturally to interpret the peak likelihood as an indication of voicing status. While a hard threshold on the peak likelihood can separate voiced and unvoiced regions reasonably-well, it is vulnerable to the random likelihood fluctuations during unvoiced regions.

In the direction of improving the robustness, instead of taking $\mathrm{max}_f \mathcal L(f, t)$, we define the peak likelihood as $\mathcal L(f_0(t), t)$ so that the voicing decision will be consistent with the F0 estimate. In addition, the peak likelihood sequence is decoded by a two-state hidden Markov model, with the states mapped to voiced/unvoiced status, to further reduce spontaneous errors. The two-state HMM requires a pair of observation distributions characterizing the peak likelihood under voiced and unvoiced frames. Following the strategy of computing the calibration function $\mathcal L_\mathrm{cal}$, yet another Monte-Carlo simulation is performed on white noise input signals, from which the peak F0 likelihood is extracted. It is empirically found that $\mathcal L(f_0(t), t)$ follows a normal distribution; on a grid-approximation of $\mathcal L(f, t)$ with 128 log-spaced frequency bins, the mean is $-4.78$ and the variance is $0.02$. Note that the mean is close to but slightly greater than the log probability mass of a uniform distribution, $\log 1/128 \approx -4.85$.

The distribution of peak likelihood on voiced regions, however, cannot be determined through simulation as the SNR of voiced speech can vary depending on the environment and linguistic context. Assuming the distribution in question is also normal, we estimate the mean and variance from the peak likelihood at run-time using Baum-Welch algorithm. The training starts with an initial mean of $-2.0$ and an initial variance of $1.0$. The transition probability between voiced and unvoiced states is fixed at $t_\mathrm{hop} / 0.2$ where $t_\mathrm{hop}$ is the time interval for F0 estimation. The binary sequence of voicing status can be efficiently estimated from the peak log likelihood using Viterbi algorithm.

\subsection{Tricks and Implementation Details}
\label{sec:tricks}

\textbf{Dithering.} It is observed that the voicing detector is prone to picking up small sinusoidal interferences during silent and unvoiced regions. A simple fix is to dither the input signal with a white noise at $2\%$ the maximal amplitude.

\noindent\textbf{Smoothing of the likelihood map.} The current design of IF and SNR estimators assumes quasi-stationary harmonic amplitude. Manual inspection of the harmonic SNR estimated from speech signals show that the amplitude modulation at vowel onsets and endings causes the SNR to be underestimated, further causing voicing decision errors at a later stage of the algorithm. To alleviate this problem, the F0 likelihood map is smoothed by a moving average filter prior to voicing detection. The order of the filter is inversely proportional to the frequency,
\begin{align}
\mathcal {\bar L}(f, t) &= \frac{f}{3} \int_{t - 1.5/f}^{t + 1.5/f} \mathcal L(f, t) dt
\end{align}

\section{Evaluation}

The proposed algorithm is evaluated on clean speech samples from CSTR \cite{bagshaw-1993} and CMU Arctic \cite{kominek-2004} databases. Objective criteria from Drugman \etal\cite{drugman-2011} are adopted to assess the accuracy of F0 and voicing status estimation. Specifically, the F0 frame error (FFE) indicates the overall performance of an estimator and it breaks down into gross pitch error (GPE) and voicing decision error (VDE). The GPE is defined as the percentage of frames whose estimated F0 deviates from the reference value by more than $20\%$, among all voiced frames with correctly estimated voicing status.

\noindent\textbf{Datasets and the ground truth.} The CSTR database contains 50 English sentences voiced by one male and one female speaker. The F0 annotations and voicing labels provided by the database are interpolated at a $5$ ms interval to be used as the reference F0 for this study. For CMU Arctic database, the first 50 sentences from two male speakers (``jmk" and ``bdl") and one female speaker (``slt") are selected; the reference F0 is extracted from EGG signals using Praat \cite{boersma-2016} with the default pitch tracking parameters also at a $5$ ms interval.

\noindent\textbf{Other methods evaluated in this test.} The following F0 and voicing estimation algorithms are compared against Nebula: YANGsaf \cite{kawahara-2016}, DIO \cite{morise-2010}, SAcC \cite{lee-2012}, RAPT \cite{talkin-1995}, Praat \cite{boersma-2016}, and SRH \cite{drugman-2011}. For all methods and all speakers, the search range for F0 is set to $[55, 400]$ Hz while all other parameters remain at their default values. The results from SAcC and SRH, only available at a $10$ ms interval, are interpolated to match the rate of the reference.

\begin{table}[h]
\centering
\begin{tabular}{|c|c|c|c|}
  \hline
  Method  & FFE$\%$ & GPE$\%$ & VDE$\%$ \\ \hline
  Nebula  & \textbf{5.53} (7.39) & \textbf{0.30} (\textbf{0.61}) & \textbf{5.39} (7.01) \\ \hline
  RAPT    & 5.96 (\textbf{6.77}) & 0.74 (1.09) & 5.61 (\textbf{6.49}) \\ \hline
  Praat   & 6.35 (8.13) & 0.57 (1.44) & 6.10 (7.78) \\ \hline
  YANGsaf & 7.33 (8.54) & 1.14 (2.56) & 6.75 (7.95) \\ \hline
  SAcC    & 7.63 (9.50) & 0.67 (1.59) & 7.29 (8.97) \\ \hline
  SRH     & 8.28 (9.82) & 0.71 (0.82) & 7.98 (9.56) \\ \hline
  DIO     & 9.08 (10.14) & 0.54 (1.07) & 8.81 (9.86) \\ \hline
\end{tabular}
\caption{\figcap{Table of average and worst-case scenario performance across all databases, ranked in ascending average FFE.}}
\vspace{-10pt}
\label{tab:evaluation}
\end{table}

Table~\ref{tab:evaluation} summarizes the results from the evaluation across all databases, including the average and worst-case\footnote{The worst-case percentage is taken across speakers, as opposed to taken across sentences.} (in parenthesis) error percentages of each algorithm. It is seen that Nebula has a clear advantage over all criteria in terms of average performance. In the worst-case scenario, while RAPT outperforms Nebula on voicing decision by a $0.5\%$ margin, Nebula still holds the second place. A major source of voicing decision errors is found to be the underestimated SNR at voiced/unvoiced boundaries, even after applying the tricks in section~\ref{sec:tricks}. It is also worth noting that Nebula reduces gross error rate to an almost negligible level ($0.3\%$). We believe that such a significant improvement over YANGsaf can be attributed to the choice of high-variance priors for model training (section~\ref{sec:priors}) and the inclusion of \{SNR2, IF2, SNR0\} features, under a carefully designed statistical framework.

\begin{figure}
  \input{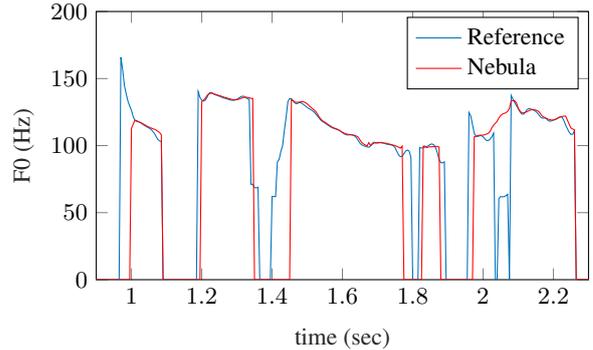}
  \caption{\figcap{An example of F0 estimated using Nebula versus the reference F0 extracted from the EGG signal, taken from the highest-error sentence on speaker ``bdl". At $1$ sec, $1.4$ sec, and $1.96$ sec, Nebula failed to track the rapid pitch rises and drops. On the other hand, at $2.05$ sec, the reference F0 is subjected to half-pitch errors. }}
  \label{fig:eval-example}
\end{figure}

\begin{table}[h]
\centering
\begin{tabular}{|l|c|c|c|c|}
  \hline
  Speaker  & FFE$\%$ & GPE$\%$ & VDE-V$\%$ & VDE-U$\%$ \\ \hline
  rl (M)   & 5.730 & 0.294 & \phantom{1}5.111 & \phantom{1}5.943 \\ \hline
  bdl (M)  & 7.386 & 0.607 & 10.742 & \phantom{1}1.030 \\ \hline
  jmk (M)  & 5.742 & 0.118 & \phantom{1}7.944 & \phantom{1}3.396 \\ \hline
  sb (F)   & 4.068 & 0.473 & \phantom{1}4.035 & \phantom{1}3.826 \\ \hline
  slt (F)  & 4.470 & 0.027 & \phantom{1}0.489 & 11.423 \\ \hline
\end{tabular}
\caption{\figcap{Breakdown analysis of the errors on each speaker.}}
\label{tab:evaluation-breakdown}
\vspace{-10pt}
\end{table}

A breakdown analysis of Nebula's performance on each speaker is shown in Table~\ref{tab:evaluation-breakdown}. The voicing decision error is divided into mis-classification rates on voiced (VDE-V) and unvoiced (VDE-U) frames. First it is seen that the algorithm performs better on female voices (``sb" and ``slt") with fewer voicing decision errors on voiced frames. Next, the largest GPE and VDE-V are observed on male speaker ``bdl" (see Figure~\ref{fig:eval-example} for a worst-case example). The large errors can be explained by the observation that ``bdl" features a less regular glottal pulse pattern compared to other speakers in the database, causing errors in both Nebula's predictions and the reference F0 (extracted from EGG signals). Finally, the large VDE-U on speaker ``slt" is also found to be caused by errors in the reference due to noises present in the EGG signals.

Concerning that the evaluation may become systematically biased due to inaccurately extracted reference F0, we repeated the analysis in Table~\ref{tab:evaluation} on speakers ``rl", ``sb" and ``jmk" only. The accuracy ranking, however, remained the same. The evaluation yields convincing evidences that if not any better, the accuracy of the proposed method is at least comparable to the state-of-the-art results on F0 estimation. A more rigorous evaluation requires expert-annotated reference F0, which has not been attempted given the limited time.

\section{Conclusions}

This paper presented Nebula, a F0 and voicing status estimation algorithm. The most significant contribution of this study is a novel methodology for speech signal analysis by characterizing the statistical properties of feature extractors using Monte-Carlo simulation (Figure~\ref{fig:overview}). The claim that the requirements on the speech prior (i.e. training data) can be relaxed for time-frequency local feature extractors is supported by an objective evaluation on 3 male and 2 female speakers: Nebula trained on fully synthetic data outperformed state-of-the-art methods on gross pitch error and achieved the overall best average performance. We believe that the statistical feature extractor modeling technique will also find applications in other topics in speech analysis, for example the estimation of spectral envelope and the decomposition of speech into periodic/aperiodic components.

\bibliographystyle{IEEEtran}

\bibliography{mybib}

\end{document}